\documentclass[twocolumn,
               showpacs,
               preprintnumbers,
               prl,
               endfloats]{revtex4}

\usepackage{amsfonts}
\usepackage{graphicx}
\usepackage{color}

\begin{document}

\title{Universal fluctuations in tropospheric radar measurements}

\author{Andrea Barucci,$^1$ Giovanni Macaluso,$^1$  Daniele Mecatti,$^1$  Linhsia Noferini,$^1$ Duccio Fanelli,$^2$ Angelo Facchini,$^3$ Massimo Materassi,$^4$ Massimiliano Pieraccini,$^1$ Carlo Atzeni,$^1$}

\affiliation{$^1$Department of Electronics and Telecommunications, University of Florence, via Santa Marta 3, 50139, Firenze, Italy, EU}

\affiliation{$^2$Dipartimento di Energetica \textquotedblleft S.Stecco \textquotedblright, University of Florence ans INFN, via S. Marta 3, 50139, Firenze, Italy, EU}

\affiliation{$^3$Department of Information Engineering and Center for the Study of Complex Systems, University of Siena, via Tommaso Pendola 37, 53100 Siena, Italy, EU}

\affiliation{$^4$Istituto dei Sistemi Complessi ISC-CNR, via Madonna del Piano 10, 50019, Sesto Fiorentino, Italy, EU}

\begin{abstract}

Radar data collected at an experimental facility arranged on purpose suggest that the footprint of atmospheric turbulence might be encoded in the radar signal statistics. Radar data probability distributions are calculated and
nicely fitted by a one parameter family of generalized Gumbel (GG) distributions, $G_{a}$. A relation
between the wind strength and the measured shape parameter $a$ is obtained. Strong wind fluctuations return pronounced asymmetric leptokurtic profiles, while Gaussian profiles are eventually recovered as the wind fluctuations decrease. Besides stressing the crucial impact of air turbulence for radar applications, we also confirm the adequacy of  $G_{a}$ statistics for highly correlated complex systems.

\end{abstract}


\maketitle

Radar sensors are currently employed in a large variety of applications, e.g. surveillance and target identification \cite{skolnik}, weather observations and forecasting \cite{atlas}, and geophysical investigations \cite{massonet}. It is therefore of paramount importance to
have a clear control on all possible sources of external disturbances,
which could significantly alter the measurements response.

In 2006, within the activities of the GALAHAD project \cite{galahad}, a large collection of ground based radar data was recorded, and soon realized to constitute a unique opportunity for investigating the impact of external perturbations on the sought response.
It was in particular observed that occasionally, but in coincidence with hard windy conditions, the radar interferograms were affected by an overall decorrelation.
Motivated by this unexpected finding, it was hypothesized that the atmospheric turbulence could drastically influence the radar propagation even over short ranges, an observation which however rested on purely qualitative ground.
No further experimental realizations were in fact subsequently developed to scrutinize the process and reach an unambiguous proof of concept.

At variance, on the theoretical side, pioneering studies on electromagnetic signal propagation through a turbulent atmosphere \cite{clifford} \cite{ishimaru} date back to the 70s. Punctual fluctuations of air refractive index  materialize in a recorded imprint, which could therefore encode an indirect signature of turbulence.
Indeed, the embedding medium can be ideally segmented in neighboring volume cells, each contributing to the signal according to an overall trend and its own statistics.
The signal impinging on the junction between two adjacent cells is partly reflected and partly transmitted and the data collected at the receiver is sensitive to the sequence of scattering events occurred along the propagation path.
In this sense, the radar echo is a global quantity indirectly representing the specific state of the crossed medium.
A comprehensive interpretative framework for such phenomena is however still lacking, following the inherent difficulties to accommodate for the intimate, highly complex nature of radar/air interactions. In particular, the characteristic of air induced fluctuations remain to be fully elucidated.

Starting from this setting, and to gain a conclusive insight into this crucial issue, we have
arranged an outdoor experiment, realized in a controlled environment.
A $k_{u}$-band radar was mounted on a plateau looking at an artificial target at fixed distance.
The location was chosen in order to reproduce as close as possible an ideal situation, in which the radar signal fluctuations would solely have depended on the propagation medium.
An anemometer, located near the target, was used to register the wind velocity.

Then, the probability distribution function of the collected signal was obtained and studied, focusing on the shape modification as function of the air wind condition.
This allowed us to return a statistical characterization of the radar fluctuations due to the atmospheric propagation, up to momenta of arbitrarily high order.
As we shall demonstrate, strong wind conditions are associated to asymmetric leptokurtic profiles,
standard Gaussian distributions being instead recovered as the wind strength decreases.
This observation can be cast in the form of a  phase transition, the wind strength acting as the external control parameter.
To establish a clear causality relation between the air turbulence, here encapsulated in the wind velocity, and the radar signal statistics constitutes the primary goal of the Letter.
Equally important, as an additional result of our analysis, we will confirm the adequacy of the so called generalized Gumbel (GG) distribution as the paradigmatic PDF of global quantities in correlated, spatially extended system \cite{bertin}.

\begin{figure}
\begin{center}
\includegraphics[width=0.8\linewidth,keepaspectratio=true]{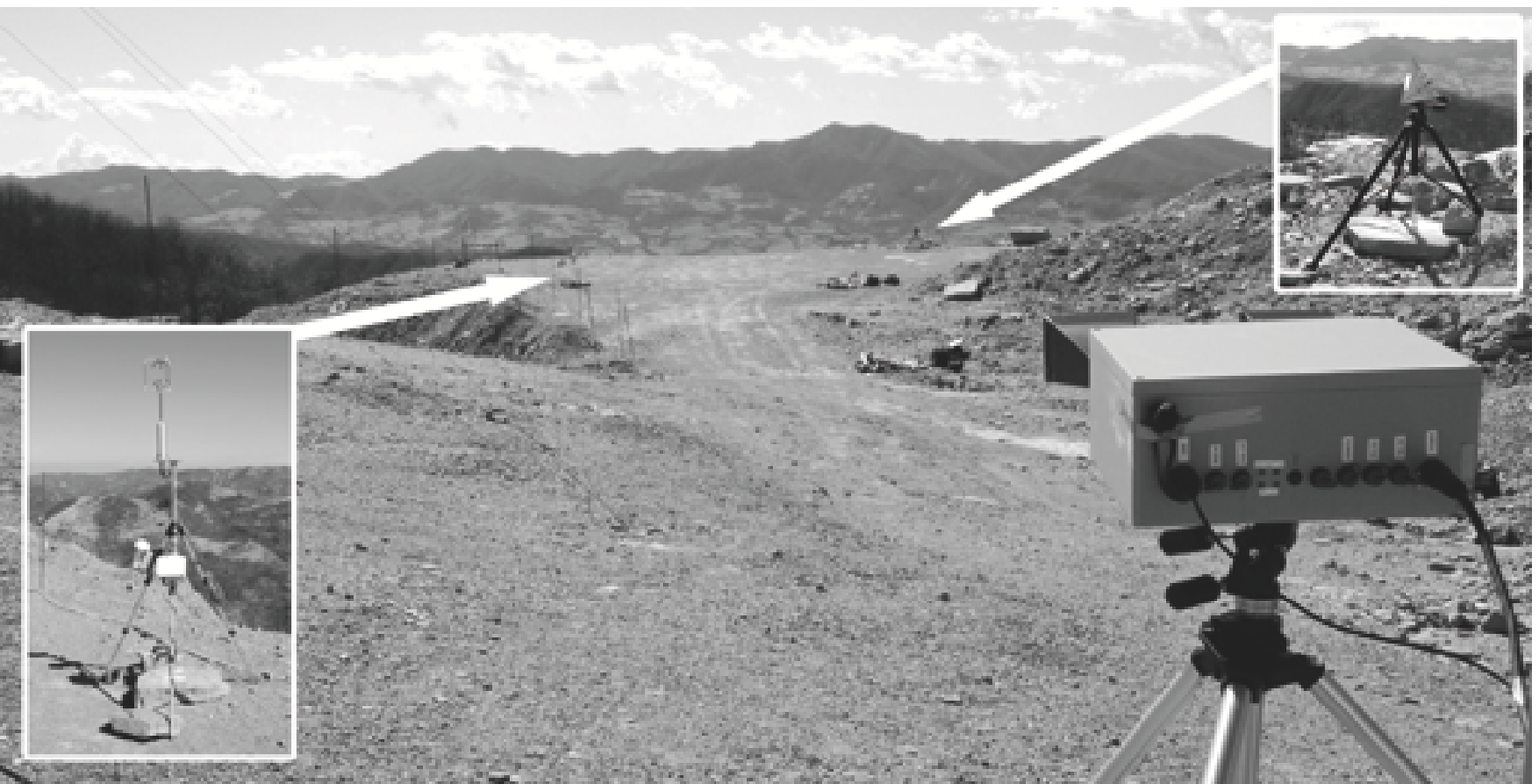}
\caption[linhsia_1]{
The radar sensor at the outdoor experimental facility. The site was equipped with a trihedral corner reflector (highlighted on the right side of the picture), posed at about 60 m from the radar, and an ultrasonic anemometer (highlighted on the left).
}
\label{fig:scene}
\end{center}
\end{figure}

In our experiment, a portable continuous wave stepped frequency $k_{u}$-band radar was installed over the excavation area of a disused quarry of Pietra Serena located near Firenzuola (Firenze, Italy)
\footnote{The data here presented were recorded from $ 22:45 $ LT on March 25, 2009, to $04:15$ LT on March 26, 2009. Clear air conditions were fulfilled during the measurement period.}.
It operated with range resolution of $1 \mathrm{\ meter}$ and unambiguous range of $600 \mathrm{\ meter}$.
Details on the radar system can be found in \cite{pieraccini}.
The observed scene was essentially a rectangular rock plateau with three sides ending with a cliff
\footnote{ It would be worth to emphasize the crucial role played by the extremely favorable conditions under which the experiment was realized. The location was chosen to minimize any other possible source of fluctuations, as those due to the vegetation. Indeed, from previous experimentation activities, we observed that vegetation may produce further unwanted scattering, blurring away any beautiful footprint of clear air turbulence in the radar signal statistics. }.
The site was equipped with an artificial radar target, a trihedral corner reflector, posed at the edge of the cliff opposite to the radar side (see Figure \ref{fig:scene}). The radar-target distance was about 60 m. An ultrasonic anemometer, working at 4 Hz sampling rate, installed near the radar, measured the wind velocity fluctuations.

Radar data were collected at a rate of about $50\mathrm{\ Hz}$, (sampling a $6$ hours long recording window). Mainly they represent the amplitude of the backscattered signal from the corner reflector after propagation to the corner and back. The data to be analyzed
were arranged in a time series $A\left( t\right) $ of about one million points.

To just focus on air's impact on the propagation, all other possible sources of disturbances need to be identified and subsequently excluded. In the current case, because of the bareness of the scene, signal fluctuations may depend on the oscillations of the radar tripod and/or the corner reflector's support, and on the intrinsic instrumental noise. Mechanical oscillations can be straightforwardly characterized in the frequency domain and their contribution to the PDF shape was found to be negligible. The instrumental noise is thermal noise, which is known to be white ($ SNR = 60 dB $). The residual signal as departing from the expected, a priori constant output, is imputed to atmospheric propagation.

Different scale processes occur within the atmosphere, generally referred to as synoptic-scale processes (dozen of days), meso-scale processes (from day to hour) and micro-scale processes (less than an hour). Turbulent phenomena concern the micro scale \cite{stull}, corresponding to distances shorter than 1 km.
In order to remove the components due to meso-scale and longer scale processes from the series $A(t)$, a multiplicative detrending was applied. The \textquotedblleft
detrended\textquotedblright\ series $A_{\det}(t)$ is determined as
\begin{equation}
A_{\det }(t) =\frac{A(t)}{A_0(t)},
\label{detrended}
\end{equation}

where $A_0\left( t\right) $ is the \textquotedblleft trend\textquotedblright\ obtained by a moving average filter, which in this case is equivalent to a low-pass filter $A\left( t\right) $ with a threshold frequency of $5.5\cdot10^{-4} \mathrm{\ Hz}$.
Once detrended, only micro-scale fluctuations are expected to contribute to the PDF.

Non-stationary modulation can still affect the detrended series, e.g., varying Reynolds regimes. A $60$ minutes long temporal window was applied sequentially, in order to analyze finite portions of the series, that are as stationary as possible, while still containing a meaningful number of samples for statistical analysis.
Each selected window consists of about $10^{5}$ points, which renders of some reliability the
study of rare fluctuations. Different windows overlapped partially, for about one third of their length.

Turbulent wind conditions associated to the different analysis windows, is quantified via the turbulent kinetic energy (TKE) indicator $K_t$ \cite{stull}. This latter is calculated from the fluctuations of the three wind velocity components $v{'}_{x}$, $v{'}_{y}$ and $v{'}_{z}$, revealed by the anemometer and reads:

\begin{equation}
K_{\mathrm{t}}=\left\langle v{'}_{x}^{2}\right\rangle
+\left\langle v{'}_{y}^{2}\right\rangle +\left\langle v{'}_{z}^{2}\right\rangle,
\label{tke}
\end{equation}

where the symbol $\langle \cdot \rangle$ stands for a time average over the inspected window.

To obtain an exhaustive representation of the signal fluctuations, beyond the second momentum, we reconstructed the
full PDF, within each selected time domain. Imagine to assume an ideal setting where interference of the radar signal with the wind can be neglected.
Then, having eliminated other possible sources of systematic errors, one expects to recover the
theoretically predicted PDF for a pointwise scatterer (the corner reflector) suffering from thermal noise.
This is the well-known Rice distribution \cite{richards}, a right skewed continuous curve,
which approaches the Gaussian for high signal-to-noise ratio. Surprisingly, and at variance with what is customarily believed, a quite different distribution appears from our data analysis: the PDF of $A_{\det }$ is an asymmetric, left skewed, distribution characterized by an exponential tail on the one side and a rapid falloff on the other.
Such a significant discrepancy can be ultimately imputed to reflect the non trivial interplay between propagating signal and turbulent dynamics of the embedding atmospheric medium, an effect which is not accommodated for in the theoretical scenario yielding to the Rice profile.

Motivated by this working ansatz, we set down to analyze more closely the experimental PDFs. To this end, and aiming at a quantitative representation of the data, we considered a non-Gaussian, one parameter family of distributions,
recently invoked to describe the fluctuations of global quantities in  turbulent context \cite{bramwell_n}.
These are the so-called generalized Gumbel (GG) distributions, originating from the study of extreme values statistics \cite{extreme_value1} \cite{extreme_value2} and more recently
proposed to universally describe the fluctuation of global quantities in correlated many degrees
of freedom systems \cite{bertin} \cite{bramwell}.
In this respect the GG  constitutes the natural extension of the Gaussian distribution which instead apply to
uncorrelated variables, as a consequence of the central limit theorem.
In the last years, GG distributions proved to adequately adapt to distinct realms of applications,
ranging  from e.g. the magnetization of the two dimensional XY model
close to the Koterlitz-Thouless transition \cite{bramwell_footnote1} \cite{bramwell_footnote2} \cite{bramwell_footnote3}
\footnote{The so GG distribution occurs, for the 2D XY model
 to an excellent approximation, along the line of critical points and it
 is exact in the limit $T \rightarrow 0$. Corrections become observable as vortex
 excitations appear near the Kosterlitz-Thouless transition.}, to the statistics of the
level of the Danube river \cite{portelli}, passing through the power injected in a turbulent flow and in electroconvetion \cite{katona}.
Complexity and high correlation should be the common feature of these systems, ending up with a GG distribution.

The GG distributions family of curves $G_{a}$ is defined as:

\begin{equation}
\left\{
\begin{array}{l}
\begin{array}{cc}
G_{a}\left( x\right) =\frac{a^{a}b\left( a\right) }{\Gamma \left( a\right) }%
\exp \left\{ a\left[ b\left( a\right) \left( x-s\left( a\right) \right)
-e^{b\left( a\right) \left( x-s\left( a\right) \right) }\right] \right\}%
\end{array}
\\
\\
\begin{array}{cc}
a\in \mathbb{R}^{+}, & b\left( a\right) =\frac{1}{\sigma _{x}}\sqrt{\frac{%
d^{2}}{da^{2}}\ln \Gamma \left( a\right) },%
\end{array}
\\
\\
s\left( a\right) =\left\langle x\right\rangle +\frac{1}{b\left( a\right) }%
\left[ \ln a-\frac{d}{da}\ln \Gamma \left( a\right) \right] .%
\end{array}%
\right.  \label{Ga}
\end{equation}%

When the real positive parameter $a$ varies from large to small values,
$G_{a}\left( x\right) $ changes its shape from that of a Gaussian distribution towards a more and more skewed PDF.
This feature of the distribution family $G_{a}$ rendered it very popular in complex dynamics, when the statistics of global quantities are dealt with.
The parameter $a$ has been proposed to be inversely related to the correlation length of the system \cite{portelli}, and it has been used for measuring the number of its effective degrees of freedom \cite{joubaud}.
As a side observation, it is worth noting here that the GG distribution with $a=\frac{\pi }{2}$ approaches the well known Bramwell-Holdsworth-Pinton distribution that was
found for the fluctuations of turbulent power consumption \cite{bramwell_n}.

\begin{figure}

\begin{center}

\includegraphics[width=1\linewidth,keepaspectratio=true]{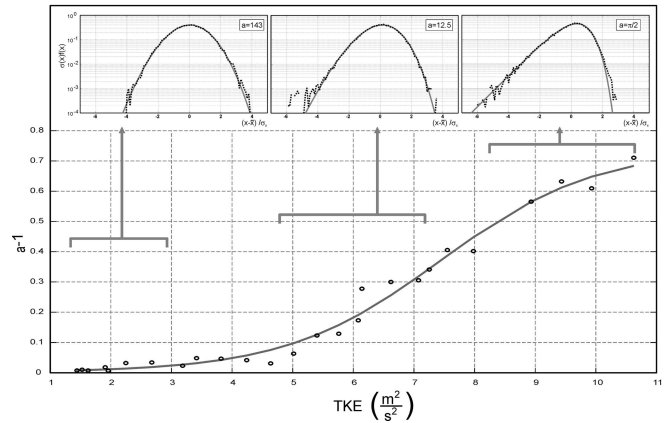}

\caption[linhsia_2]{
The PDF of the series $ A_{\det}\left(t\right) $ (dot line) and the best-fit GG distribution (solid line) shown in the upper panels refer to three different analysis windows. The value of the shape parameter $a$ as resulted from the fitting procedure is also reported on each panel
(notice that $G_{a}$ depends on the gamma function, which limits the maximum accessible value for the shape parameter to $a=143$).
The shape of the distribution varies from Gaussian, for low TKE values (upper panel, left), to Bramwell-Holdsworth-Pinton distribution \cite{bramwell_n}, for high TKE values (upper panel, right). The general agreement confirms the adequacy of the GG distribution for describing the radar data fluctuations.\\
Distributions with shape similar to the ones reported in the panels are found for TKE values within the intervals highlighted below the panels. This suggests a relationship between the shape, and hence the $a$ parameter values, and the wind fluctuations intensity. The scatter plot in the main panel of the figure reports the inverse of the shape parameter, $a^{-1}$, vs. TKE. The direct driving of the turbulent wind in the medium toward asymmetric statistics, is stressed by the solid line representing a fit based on the sigmoid profile:
$ \lambda_1 + \lambda_2 \tanh\left[ \lambda_3(x-\lambda_4) \right]$
with $ \lambda_i $ free parameters.
}

\label{fig:trans}

\end{center}

\end{figure}

Back to the data collected in our experiment, properly rescaled fluctuations are calculated and the normalized experimental histograms reconstructed. These latter are then interpolated via $G_{a}\left( x\right)$, $a$ being adjusted as the best fit parameter. In Figure \ref{fig:trans} three representative curves are displayed which testified
on the adequacy of the proposed ansatz. The general agreement between data and GG distributions is remarkable.
It's worth noting that we didn't find any correspondence between the statistical behavior of the radar signal and TKE, approaching the distribution of TKE a Rayleigh distribution on every analyzed window.

Even more interestingly, the values of $a$ as obtained from the fitting procedure appear to be strongly correlated to the quantity $K_{\mathrm{t}}$, representing the
turbulent wind intensity within the crossed medium, for every chosen (statistically stationary) data segment. This finding, reported in the main panel of
Figure \ref{fig:trans}, materializes in a genuine transition for $a^{-1}$ vs. $K_{\mathrm{t}}$, showing the direct footprint of turbulence in the medium
on the non-Gaussian nature of the radar fluctuations.

The transition between Gaussian and a non-Gaussian regime is rather sharp and takes place for a specific value of
$K_{\mathrm{t}} = K_{\mathrm{t}}^{c}$ ($\simeq 6\mathrm{\ m}^{2}\mathrm{s}^{-2}$, in this case).
The robustness of the scenario here depicted is also tested versus modulating the size of the windows, in which the full time series is being decoupled: the transition point is not smeared over a wider $K_{\mathrm{t}}$ interval, but conversely, it remains strictly positioned in $K_{\mathrm{t}}^{c}$.
Notice however, that the value of $K_{\mathrm{t}}^{c}$ was obtained keeping fixed the geometry of the experiment, in particular the radar-target distance: it would be interesting to vary this distance and monitor the consequent changes on the estimated value of $K_{\mathrm{t}}^c$.
An experimental study on this specific point is already being planned, and results will be reported in a further publication.
These latter results entail the possibility of establishing a control flag in wind intensity, when the device is being operated in weather conditions which are likely to strongly influence the data acquisition process.
Indeed, measurements fluctuating according to skewed distributions could in principle induced biased conclusions (e.g. position of the mean) without proper information.
Moreover, it can be speculated that the fitting parameter $a$ returns an indirect measure of the correlation degree among air volume cells, in which the emitter-detector distance
can be ideally segmented. When the wind fluctuations are pronounced,
correlations become certainly crucial over the finite emitter-detector separation distance and $a \rightarrow \pi/2$, as seen in
\cite{Portelli1}. Qualitative agreement is also found in the opposite limit, when the wind turns weak.

Summing up, in this study it is demonstrated that the fluctuations in time of a radar signal propagating through the atmosphere, even over short ranges (less than $100$ $\mathrm{m}$) and moderate wind conditions, are generally distributed according to non Gaussian PDFs. Moreover, they generally agree with the GG distributions described in (\ref{Ga}). These distributions, have been recently shown to describe very well the fluctuations of global quantities in correlated systems, turbulent systems and many other natural phenomena. The conjecture of \textit{wind driven fluctuations} in the radar measurements is further strengthen by the observation that the shape parameter $a$ of the GG distribution fitting the radar data PDFs is directly related to the TKE of the crossed medium.
All these facts clearly support the idea that the non-Gaussian fluctuations of radar data are determined by the wind strength. The developing of a consistent model for radar signal transmission through a windy atmosphere, according to
 which the GG distribution emerges from the physical mechanisms of scattering, will be the focus of forthcoming works.


\end{document}